\documentclass[superscriptaddress,showpacs,amssymb,10pt,aps,prd,reprint,longbibliography,footinbib]{revtex4-1}
\usepackage{graphicx,epsfig,amssymb,times} 
\usepackage{amsmath,amsfonts}
\usepackage{bm}
\usepackage{epstopdf}
\usepackage[linktocpage,colorlinks]{hyperref}
\usepackage[caption=false]{subfig}
\usepackage[usenames]{color}     
\usepackage{float}
\usepackage{natbib}
\usepackage{ulem}
\usepackage{cancel}
\usepackage{verbatim}
\usepackage{enumitem}
\usepackage[utf8]{inputenc}
\definecolor{coolblack}{rgb}{0.0, 0.18, 0.39}
\definecolor{darkred}{rgb}{0.5,0,0}
\definecolor{darkgreen}{rgb}{0,0.5,0}
\definecolor{darkblue}{rgb}{0,0,0.5}
\definecolor{lapislazuli}{rgb}{0.15, 0.38, 0.61}
\definecolor{venetianred}{rgb}{0.78, 0.03, 0.08}
\definecolor{bleudefrance}{rgb}{0.19, 0.55, 0.91}
\definecolor{dogwoodrose}{rgb}{0.84, 0.09, 0.41}
\hypersetup{colorlinks=true, citecolor=darkblue, linkcolor=darkblue, 
urlcolor = darkblue}

\newcommand{\nn}{\nonumber\\}
\newcommand{\ep}{\epsilon}

\usepackage{amsmath}
\usepackage{centernot}
\newcommand{\be}{\begin{equation}}
\newcommand{\bw}{\begin{widetext}}
\newcommand{\ew}{\end{widetext}}
\newcommand{\bse}{\begin{subequation}}
\newcommand{\ese}{\end{subequation}}
\newcommand{\ee}{\end{equation}} 
\newcommand{\eei}{\end{eqnarray}\indent\indent}
\newcommand{\bc}{\begin{center}}
\newcommand{\ec}{\end{center}}
\newcommand{\ber}{\begin{eqnarray}}
\newcommand{\eer}{\end{eqnarray}}
\newcommand{\ba}{\begin{array}}
\newcommand{\ea}{\end{array}}
\newcommand{\bal}{\begin{align}}
\newcommand{\eal}{\end{align}}

\newcommand{\sfrac}[2]{{\textstyle{#1\over#2}}}
\def\case#1/#2{\textstyle\frac{#1}{#2} }
\newcommand{\nb}{\nabla}
\newcommand{\D}{\tl\nb}

\newcommand{\tl}{\tilde}
\newcommand{\bver}{\begin{verbatim}}
\newcommand{\ever}{\end{verbatim}}

\begin{document}

\title{\large Shear-free conditions of a Chaplygin-gas-dominated universe}

\author{Amare Abebe}\email{Amare.Abbebe@gmail.com}
\affiliation{Center for Space Research, North-West University, Mahikeng 2745, South Africa}

\author{Mudhahir Al Ajmi}\email{Mudhahir@gmail.com}
\affiliation{Department of Physics, College of Science, Sultan Qaboos University,
P.O. Box 36, Al-Khodh 123, Muscat, Sultanate of Oman}

\author{Maye Elmardi}\email{Maye.Elmardi@gmail.com}
\affiliation{Center for Space Research, North-West University, Potchefstroom 2520, South Africa}

\author{Hemwati Nandan}\email{Hnandan@associates.iucaa.in}
\affiliation{Department of Physics, Gurukula Kangri Vishwavidyalaya, Haridwar - 249407, India}
\affiliation{Center for Space Research, North-West University, Mahikeng 2745, South Africa}
\author{Noor ul Sabah}\email{Sabah.Noon@gmail.com }
\affiliation{Department of Physics, College of Science, Sultan Qaboos University,
P.O. Box 36, Al-Khodh 123, Muscat, Sultanate of Oman}
\date{\today}

\begin{abstract}
In this work, we revisit the shear-free conjecture of general relativity and show the violation of the well-known shear-free condition for perfect-fluid spacetimes. It had been shown in previous investigations that, in the general relativistic framework, the matter congruences of shear-free perfect fluid spacetimes should be either expansion-free or rotation-free. Our current investigation, however, indicates that a universe dominated by a Chaplygin gas can allow a simultaneous expansion and rotation of the fluid provided that certain non-trivial conditions, which we derive and describe below, are met. We also show that, in the appropriate limiting cases, our results reduce to the expected results of dust spacetimes which can only expand or rotate, but not both, at the same time.
\end{abstract}

\maketitle
\section{Introduction}

One of the most intriguing results obtained in general relativity (GR) is obtained by G\"{o}del, who showed \cite{godel52} that shear-free time-like geodesics  of spatially homogeneous universes cannot expand and rotate simultaneously. This observation was later generalized by Ellis \cite{ellis67} to include inhomogeneous cases of shear-free time-like geodesics. 
This result remains true for all linearized cases for all physically realistic barotropic perfect fluids \cite{nzioki2011shear}, except for the case of a specific highly non-linear equation of state (EoS) which is considered non-physical. 
However, there are no such restrictions in Newtonian gravitation theory \cite{ellis2011,narlikar99,narlikar63}. This is therefore problematic because Newtonian theory is a limiting case of GR. Even more interesting is that, in the linearized fourth-order gravity theory, there are solutions for which the universe can both rotate and expand simultaneously having the same properties as of Newtonian theory, but not of GR \cite{abebe2011shear}. 

Geodesics play an important role in the description of fluid flows in astrophysics and cosmology. The kinematics used of such fluid flows are described by three kinematical quantities: the expansion $\Theta$, shear (or distortion) $\sigma_{ab}$ and rotation $\omega^c$, along with acceleration $A_a$ of the four-velocity field $u^a$ tangent to the fluid flow lines. The governing equations are obtained by contracting the Ricci identities (applied to $u^a$) along and orthogonal to $u^a$, and determine how they couple to gravity \cite{EllisCovariant, nandan09, nandan15}.

The Chaplygin Gas (CG) model in cosmology is one of the widely explored candidates for the dark matter-dark energy conundrum. It may thus be considered as alternative for the FLRW universe model \cite{chap1904} with an exotic perfect-fluid EoS \cite{gorini2003,bento02,bento03,bilic02,deb04,dev03} able to describe a smooth transition between an inflationary phase, the matter-dominated decelerating era, and then a late-time accelerated de Sitter phase of cosmic expansion \cite{kamen01,gorini03,fabris02,bean03, bouh15,tup, pas}.

The main objective of this work is to study the existence of simultaneously expanding and rotating shear-free spacetimes described by linearized perturbations of a CG dominated universe in a fully covariant way.

In this paper we use the natural units ($\hbar=c=k_{B}=8\pi G=1$) throughout. The Latin indices run from 0 to 3.
The symbols $\nabla$, $\D$ and the overdot ($^{.}$) represent respectively, the usual covariant derivative, the spatial covariant derivative, and differentiation with respect to cosmic time. We use the
$(-+++)$ spacetime signature.

\section{The Field Equations of a Chaplygin-gas Universe}
The standard Einstein-Hilbert action for GR is given by
\begin{equation}\label{lagfR}
\mathcal{A}=\frac{1}{2}\int {\rm d}^4 x \sqrt{-g}\left[R+2{\cal L}_{m}\right]\;,
\end{equation}
where $R$ is the Ricci scalar and $\mathcal{L}_m$ corresponds to 
the matter Lagrangian.
For models of gravitation with the above action, applying the variational principle--varying the action with respect to the metric, $g_{ab}$--gives rise to the following field equations:
\be\label{efes}
G_{ab}= T_{ab}\;,
\ee
where  $G_{ab}\equiv R_{ab}-\frac{1}{2}Rg_{ab}$  is the Einstein tensor, a covariantly conserved quantity, and  $T_{ab}$ is the total energy momentum tensor (EMT), similarly conserved,  given by:
\be
T_{ab} = \mu u_{a}u_{b} + ph_{ab}+ q_{a}u_{b}+ q_{b}u_{a}+\pi_{ab}\;.
\ee 
Here $u^a$ is the direction of a timelike observer, 
$h^a_b=g^a_b+u^au_b$ is the projected metric on the 3-space perpendicular to $u^a$.  
Also $\mu$, $p$, $q_a$ and $\pi_{ab}$ denote the standard matter thermodynamic quantities: density, 
isotropic pressure, heat flux and anisotropic stress, respectively. 

CG fluid models are perfect-fluid models currently posing as possible candidates to unify dark energy and dark matter. These fluid models were originally studied in \cite{chap1904} in the context of aerodynamics, but only recently did they see cosmological applications \cite{gorini2003,bento02,bento03,fabris02,bouh15}. Among the interesting features of these models is that in the FLRW framework, a smooth transition between an inflationary  phase, the matter-dominated decelerating era, and then late-time accelerated de Sitter phase of cosmic expansion can be achieved \cite{bilic02,kamen01}.
In the original idea, the negative pressure associated with the CG  models is related to the (positive) energy density through the EoS 
\be\label{cgeos}
p=-\frac{A}{\mu^\alpha}\;,
\ee
for positive constant $A$ and $\alpha=1$. But this was later generalized to include other values of $\alpha$ such that $0 \leq \alpha \leq 1$(generalized Chaplygin gas) or to include modifications to the form of the EoS itself (modified Chaplygin gas, the generalized cosmic Chaplygin gas, etc) \cite{kamen01,elmardi17} .
For $A=0$, the cosmology based on this model reduces to that of dust (pressureless matter).
One of the  first cosmological interpretations of such a fluid model was given in \cite{barr87} where for flat universes, Eq. \eqref{cgeos} corresponds to a viscosity term that is  inversely proportional to dust energy density. Ever since the discovery of cosmic acceleration, however, both the original and generalized Chaplygin gas models have been extensively investigated as alternatives to dark energy and/or unified dark energy and dark matter models (see, e.g., \cite{bento02,bilic02,kamen01,bean03,bouh15,mak03,set07}).
\section{Linearized Field equations about FLRW background}

In the covariant description of spacetime (a somewhat detailed review of which can be found in \cite{abebe13, ellis98}), the average motion of a cosmological fluid at a point can always be represented by a family of preferred worldlines in spacetime, with a uniquely defined average 4-velocity with respect to fundamental observers associated with the worldlines given as 
\be\label{ua}
u^{a}=\frac{dx^{a}}{d\tau},~~~~~~~~~~~u_{a}u^{a}=-1,
\ee
where $\tau$ is the proper time measured along the worldlines. For any $u^{a}$, there exist unique projection tensors with the following properties:
\begin{align}
&U^{a}{}_{b}=-u^{a}u_{b}\Rightarrow U^{a}{}_{c}U^{c}{}_{b}=U^{a}{}_{b},~U^{a}{}_{a}=1,~U_{ab}u^{b}=u_{a}\\
&\label{hab}h_{ab}=g_{ab}+u_{a}u_{b}\Rightarrow h^{a}{}_{c}h^{c}{}_{b}=h^{a}{}_{b},~h^{a}{}_{a}=3,h_{ab}u^{b}=0.
\end{align}
$U^{a}{}_{b}$ projects \textit{along} the 4-velocity vector $u^{a}$ whereas $h_{ab}$ projects  the metric properties of the instantaneous restspaces of observers \textit{orthogonal} to $u^{a}$. This spacetime-splitting mechanism naturally defines the 3-volume element as
\be
\ep_{a b c}=-\sqrt{|g|}\delta^0_{\left[ a \right. }\delta^1_b\delta^2_c\delta^3_{\left. d \right] }u^d\;,
\label{eps1}
\ee
with the following identities,
\be
\ep_{a b c}\ep^{d e f}=3!h^d_{\left[ a \right. }h^e_bh^f_{\left. c \right] }\;,
\ep_{a b c}\ep^{d e c}=2!h^d_{\left[ a \right. }h^e_{\left. b \right] } \;.
\label{eps2}
\ee

Since it is a time-space split formalism, we define a covariant time derivative \textit{along} the fundamental worldlines
\be\label{convectived}
\dot{T}^{a...b}{}_{c...d}\equiv u^{e}\nb_{e}T^{a...b}{}_{c...d}\;,
\ee
and a fully \textit{orthogonally} projected covariant derivative
\be\label{spatiald}
\tl\nb_{e}T^{a...b}{}_{c...d}\equiv h^{f}{}_{e}h^{a}{}_{g}...h^{b}{}_{i}h^{t}{}_{c}...h^{m}{}_{d}\nb_{f}T^{g...i}{}_{t...m}
\ee
for any tensor $T^{a...b}{}_{c...d}$. 
Orthogonal projections of vectors, the orthogonally projected symmetric trace-free (PSTF)  part of tensors, and orthogonal projections of covariant time derivatives along $u^{a}$ (known as \textit{`Fermi derivatives'}) are denoted by angular brackets as follows:
\begin{align}\label{PSTF}
&v^{\langle a\rangle}=h^{a}{}_{b}v^{b},~~~~T^{\langle ab\rangle}=\left[h^{(a}{}_{c}h^{b)}{}_{d}-\frac{1}{3}h^{ab}h_{cd}\right]T^{cd},\\
&\dot{v}^{\langle a\rangle}=h^{a}{}_{b}\dot{v}^{b},~~~~\dot{T}^{\langle ab\rangle}=\left[h^{(a}{}_{c}h^{b)}{}_{d}-\frac{1}{3}h^{ab}h_{cd}\right]\dot{T}^{cd}.
\end{align}

The covariant derivative of the timelike vector $u^a$ can now be decomposed into the
irreducible parts as
\be
\nb_au_b=-A_au_b+\frac13h_{ab}\Theta+\sigma_{ab}+\ep_{a b c}\omega^c\;,
\ee
where $A_a=\dot{u}_a$ is the acceleration, $\Theta=\tl\nb_au^a$ is the expansion, 
$\sigma_{ab}=\tl\nb_{\langle a}u_{b \rangle}$ is the shear tensor and $\omega^a=\ep^{a b c}\tl\nb_bu_c$ 
is the vorticity vector. Similarly the Weyl curvature tensor can be decomposed 
irreducibly into the gravito-electric and gravito-magnetic parts as,
\be
E_{ab}=C_{abcd}u^cu^d=E_{\langle ab\rangle}\;;\; H_{ab}=\frac12\ep_{acd}C^{cd}_{be}u^e=H_{\langle ab\rangle}\;,
\ee
which provides a covariant description of tidal forces and gravitational radiation
respectively.\\
\subsection{Background thermodynamics and evolution}
 
Considering shear-free perturbations, the shear tensor $(\sigma_{ab})$ and its derivatives vanish identically and the linearised field equations are then given as:
 
 \begin{itemize}
\item {\bf Ricci Identities}:
In terms of the following identities
\be
\dot{\Theta}-\tl\nb_aA^a=-\frac13 \Theta^2-\frac12(\mu+3p)\;,
\label{R1}
\ee
\be
(C_0)^{ a b}:=E^{a b}-\tl\nb^{\langle a}A^{b \rangle}-\sfrac{1}{2}\pi^{ab}=0\;,
\label{R2}
\ee
\be 
\dot{\omega}^{\langle a \rangle}-\sfrac{1}{2}\ep^{abc}\tl\nb_bA_c=-\frac23\Theta\omega^a\;,
\label{R3}
\ee
\be
(C_1)^a:=\tl\nb^a\Theta-\sfrac{3}{2}\ep^{abc}\tl\nb_b\omega_c-\sfrac{3}{2}q^{a}=0\;,
\label{R4}
\ee
\be 
(C_2):=\tl\nb^a\omega_a=0\;,
\label{R5}
\ee
\be
(C_3)^{ a b}:=H^{a b}+\tl\nb^{\langle a}\omega^{b \rangle}=0\;.
\label{R6}
\ee
\item {\bf (Contracted) Second Bianchi Identities}:
\be
\dot{E}^{\langle a b\rangle}-\ep^{cd\langle a}\tl\nb_cH^{\rangle b}_d=-\Theta E^{ab}-\sfrac{1}{2}\dot{\pi}^{ab}-\sfrac{1}{2}\tl\nb^{\langle a}q^{b\rangle}-\sfrac{1}{6}\Theta\pi^{ab}\;,
\label{B1}
\ee
\be
\dot{H}^{\langle ab \rangle}+\ep^{cd\langle a}\tl\nb_cE^{\rangle b}_d=-\Theta H^{ab}+\sfrac{1}{2}\ep^{cd\langle a}\tl\nb_c\pi^{\rangle b}_{d}
\label{B2}\;,
\ee
\be
(C_4)^a:=\tl\nb^ap +(\mu+p) A^a=0\;,
\label{B3}
\ee
\be\label{B4}
\dot{\mu}=-(\mu+p)\Theta\;,
\ee
\be
(C_5)^a:=\tl\nb_bE^{a b}+\sfrac{1}{2}\tl\nb_{b}\pi^{ab}-\frac13\tl\nb^a\mu+\frac13\Theta q^{a}=0\;,
\label{B5}
\ee
\be
(C_6)^a:=\tl\nb_bH^{a b}+(\mu+p)\omega^a+\frac12\ep^{abc}\tl\nb_{b}q_{c}=0\;.
\label{B6}
\ee
\end{itemize} 
Here it is worth noting that the constraints $(C_1)^a$, $(C_2)$, $(C_3)^{a b}$, $(C_5)^a$ and $(C_6)^a$ 
are the constraints of the full Einstein field equations for general matter and are shown 
to be consistently {\it time propagated} along $u^a$ locally. However, the conditions 
$\pi_{a b}=0$  and $q^a=0$ give rise to the two new constraints $(C_0)^{a b}$ and $(C_4)^a$ 
respectively. 

We also use the following linearized commutation relations for shear-free congruences. For any scalar `$\phi$',
\ber
[\tl\nb_a\tl\nb_b-\tl\nb_b\tl\nb_a]\phi&=&2\ep_{a b c}\omega^c\dot \phi \;, \nonumber\\ 
\ep^{a b c}\tl\nb_b\tl\nb_c \phi&=&2\omega^a \dot \phi\;.
\label{C1}
\eer
If the gradient of the scalar is of the first order, we then have 
\ber
[\tl\nb^a\tl\nb_b\tl\nb_a-\tl\nb_b\tl\nb^2]\phi&=&\sfrac{1}{3}\tl{R}\tl\nb_{b}\phi\;,
\label{C2}
\eer
\ber
[\tl\nb^2\tl\nb_b-\tl\nb_b\tl\nb^2]\phi&=&\sfrac{1}{3}\tl{R}\tl\nb_{b}\phi+2\ep_{dbc}\tl\nb^d(\omega^c\dot \phi)
\label{C3}\;,
\eer
where $\tl{R}=2\left(\mu-\frac13\Theta^2\right)$ is the 3-curvature scalar.
Also for any first order 3-vector $V^a=V^{\langle a \rangle}$, we have
\ber
[\tl\nb^a\tl\nb_b-\tl\nb_b\tl\nb^a]V_a&=&\frac13\tl{R}h^a{}_{\left[a\right.}
V_{\left. b \right]}\;,
\label{C4}
\eer
\ber
h^{a}{}_{c}h^{d}{}_{b}(\tl\nb_dV^c)\dot{}=\tl\nb_b\dot{V}^{\langle a \rangle}-\sfrac{1}{3}\Theta \tl\nb_bV^a\;,
\label{C6}
\eer
\ber
h^{a}{}_{c}(\tl\nb^2V^c)\dot{}=\tl\nb_b(\tl\nb^{\langle b}V^{a \rangle})\dot{}-\sfrac{1}{3}\Theta \tl\nb^2V^a
\label{C5}\;.
\eer
Using the field equations and identities of this section, we will now investigate the 
compatibility of the new constraints with the existing ones 
in terms of the consistency of their spatial and temporal propagations. 

\section{Constraints of the shear-free condition}
We have already seen that the conditions of shear-free perturbations and the matter 
being a perfect fluid in the perturbed spacetime gives the new constraints 
$(C_0)^{a b}$ and $(C_4)^a$ respectively, with $\pi^{ab}=0=q^a$. To check their compatibility with the 
existing constraints of Einstein field equations, we plug $(C_0)_{b d}$ 
in $(C_5)_b$ which leads
\be 
\tl\nb^d\tl\nb_{\langle b}A_{d \rangle}-\sfrac{1}{3}\tl\nb_b\mu=0\;.
\label{subs1}
\ee
From the constraint $(C_4)_b$ and using the CG EoS Eq. \eqref{cgeos}, we have 
\be\label{subs2} 
A_b=-\frac{A}{\mu^2(\mu-A/\mu)}\tl\nb_b\mu=-A\D_b\phi\;,
\ee
where 
\be
\phi=\frac{1}{2A}\ln(1-A/\mu^2)\;.
\ee
We note that
\be\label{evphi}
\dot{\phi}=-\frac{1}{\mu^2}\Theta\;.
\ee
Using Eq. \eqref{subs2} in \eqref{subs1}, we obtain a new constraint equation, to linear order, given by
\be\label{nc}
A\tl\nb^d\tl\nb_{\langle b}\tl\nb_{d \rangle}\phi+\sfrac{1}{3}\tl\nb_b\mu=0\;.
\ee
The field equations must satisfy this new constraint, which, in turn must, consistently propagate along spatial and temporal hypersections. To check for spatial consistency, we take the curl of Eq. \eqref{nc} such that,
\ber\label{cnc}
&&A\ep^{acb}\tl\nb_c\tl\nb^d\tl\nb_{\langle b}\tl\nb_{d \rangle}\phi+\sfrac{1}{3}\ep^{acb}\tl\nb_c\tl\nb_b\mu=0\;.
\eer
Breaking the PSTF part according to equation Eq.  \eqref{PSTF} and using the commutators
\eqref{C2} and \eqref{C3} together with Eq. \eqref{C1}, and keeping terms only up to the linear order, we thus have
\ber\label{36}
&&A\ep^{acb}\left[\sfrac{2}{3}\tl\nb_c\tl\nb_b\tl\nb^2\phi+\sfrac{1}{3}\tl{R}\tl\nb_c\tl\nb_b\phi+\dot{\phi}\ep_{dbk}\tl\nb_c\tl\nb^d\omega^k\right]+\sfrac{2}{3}\omega^a\dot{\mu}=0\;.\nn
\eer
Again using Eqs. \eqref{C1} and \eqref{eps2} in Eq. \eqref{36} and linearizing it,
we have
\ber\label{toberearr}
&&A\left[\sfrac{2}{3}\tl{R}\omega^a\dot{\phi}-\dot{\phi}\tl\nb_k\tl\nb^a\omega^k+ \dot{\phi}\tl\nb^2\omega^a\right]+\sfrac{2}{3}\omega^a\dot{\mu}=0\;.
\eer
Now from the relation Eq. \eqref{C3} and using Eq. \eqref{R5}, we know that
\be
\tl\nb_k\tl\nb^a\omega^k=\sfrac{1}{3}\tl{R}\omega^a\;,
\ee
and using it and Eq. \eqref{evphi}, we can recast Eq. \eqref{toberearr} as,

\ber\label{scc}
&&\left[\frac{A\tl{R}}{3\mu^2}+\frac{2}{3}(\mu+p)\right]\Theta\omega^a+\frac{A\Theta}{\mu^2}\tl\nb^2\omega^a=0\;.
\eer
Here, one can easly notice that for $A=0$, i.e., dust, the equation reduces to the well-known condition in GR $i.e.$,
\be
\mu\Theta\omega^a=0\implies\Theta\omega^a=0\;.
\ee
This means that shear-free geodesics of the matter congruence in the perturbed spacetime should be either expansion-free or vorticity-free, or may be both. For the CG model, we have,
\be
\Theta\left[\frac{2}{3}\left(\mu-\frac{A\Theta^2}{3\mu^2}\right)\omega^a+\frac{A}{\mu^2}\tl\nb^2\omega^a\right]=0\;,
\ee
which indicates that either the matter congruence should be expansion-free or the vorticity-free must satisfy the condition
\be
\frac{2}{3}\left(\mu-\frac{A\Theta^2}{3\mu^2}\right)\omega^a+\frac{A}{\mu^2}\tl\nb^2\omega^a=0\;.
\ee
Here we have used the fact that $\tl{R}=2(\mu-1/3\Theta^2)$. 

In order to check for the temporal consistency, we take the time derivative of Eq. \eqref{scc}:
\be\label{tc}
\left[\frac{2}{3}\left(\mu-\frac{A\Theta^2}{3\mu^2}\right)\Theta\omega^a\right]^{.}+\left(\frac{A\Theta}{\mu^2}\tl\nb^2\omega^a\right)^{.}=0\;.
\ee
Now, using the following evolution relations
\ber
&&\dot{\mu}=\frac{A-\mu^2}{\mu}\Theta\;,\\
&&\dot{\Theta}=-\frac{1}{3}\Theta^2+\frac{3A-\mu^2}{2\mu}-A\D^2\phi\;,\\
&&\dot{\omega}^{a}=\left(\frac{A}{\mu^2}-\frac{2}{3}\right)\Theta\omega^{a}\;,\\
&&\D_b\left(\D^{\langle b}\omega^{a\rangle}\right)^{.}=-\frac{\Theta}{2}\left(1-\frac{A}{\mu^2}\right)\left(\D^2\omega^a+\D_b\D^a\omega^b\right)\;,\nn\\
&&(\D^{2}\omega^a)^{.}=\left(\frac{A}{2\mu^2}-\frac{5}{6}\right)\Theta\D^{2}\omega^a+\frac{(A-\mu^2)}{6\mu^2}\Theta\tl{R}\omega^a\;,\\
&&\dot{\tl{R}}=-2\Theta\left(\frac{2}{3}\mu-\frac{2}{9}\Theta^2+\frac{2}{3}A\D^2\phi\right)\;
\eer
Eq. \eqref{tc} can be expanded to linear order as,
\ber\label{s-fcg}
&&\Bigg\{6\mu^5\left(\mu+4\Theta^2\right)+3A^2\left(6\mu^2-\frac{3}{2}\tl{R}^2\right)\nn
&&+A\mu^2\left[\tl{R}\left(3\mu+\Theta^2\right)-24\mu\left(\mu+\Theta^2\right)\right]\Bigg\}\omega^a\nn
&&-3A\Bigg[9A\left(\mu-\Theta^2\right)-\mu^2\left(3\mu-5\Theta^2\right)\Bigg]\D^2\omega^a=0\;.
\eer
This equation automatically reduces to the well-known shear-free dust result for the case $A=0$:
\be\label{dusteq}
\omega^a(\mu+4\Theta^2)=0\;.
\ee
Now since the term inside the brackets is always positive, $\omega^a=0$ is the only solution that satisfies this equation.  In other words, we can have both $\omega^a=0$ and $\Theta=0$ in Eq. \eqref{dusteq}, leading to a vorticity-free and non-expanding fluid model, or we can have $\Theta\neq 0$ and $\omega^a=0$ (expanding but non-rotating), but we cannot have  it with $\Theta\neq 0$ and $\omega^a\neq 0$ (simultaneously expanding and rotating) as per the structure of this equation. \\

For the general case of Eq. \eqref{s-fcg} with $A\neq 0$, all we can claim at this stage is that ${\omega^a}\neq 0$ does not automatically guarantee ${\Theta= 0}$ and also $\Theta\neq 0$ does not guarantee that $\omega^a=0$ for Eq. \eqref{s-fcg} to be satisfied.

To further simplify our calculations a bit further--in an effort to find some concrete counter examples to the well-known dust results of the shear-free conjecture-- we employ the Laplace-Beltrami operator with an eigenvalue $-\lambda$ such that
\be
\D^2\omega^a=-\lambda\omega^a\;,
\ee
and rewrite Eq. \eqref{s-fcg} as
\ber
&&\left[6\mu^5(\mu+4\Theta^2)+3A^2(6\mu^2-\frac{3}{2}\tl{R}^2)\right.\nn
&&\left.+A\mu^2\Bigg(\tl{R}(3\mu+\Theta^2)-24\mu(\mu+\Theta^2)\Bigg)\right.\nn
&&\left.+3A\lambda\Bigg(9A(\mu-\Theta^2)-\mu^2(3\mu-5\Theta^2)\Bigg)\right]\omega^a=0\;.
\eer
One can see from this equation that $\omega^a$ does not have to vanish provided that expansion $\Theta$ satisfies the equation
\ber
&&\left[6\mu^5(\mu+4\Theta^2)+3A^2(6\mu^2-\frac{3}{2}\tl{R}^2)\right.\nn
&&\left.+A\mu^2\Bigg(\tl{R}(3\mu+\Theta^2)-24\mu(\mu+\Theta^2)\Bigg)\right.\nn
&&\left.+3A\lambda\Bigg(9A(\mu-\Theta^2)-\mu^2(3\mu-5\Theta^2)\Bigg)\right]=0\;.
\eer
it can be noticed this is possible provided that

\bw
\be
\Theta=\pm\sqrt{\frac{A^2\left(9\tl{R}^2/2-18\mu^2-27\lambda\mu\right)-3A\mu^3\left(\tl{R}-8\mu-3\lambda\right)-6\mu^6}{A\left(\tl{R}\mu^2-24\mu^3+15\lambda\mu^2-27A\lambda\right)+24\mu^5}}\;.
\ee
\ew

Some special cases for which simultaneously expanding and rotating solutions can exist.
\begin{enumerate}[label=(\roman*)]
\item For flat space, $\tl{R}=0\;,\lambda\neq 0$

\be
\Theta=\pm\sqrt{\frac{\mu\left(\mu^2-3A\right)\left[2\mu^3-A(2\mu+3\lambda)\right]}{A\left[9A\lambda+\mu^2(8\mu-5\lambda)\right]-8\mu^5}}\;,
\ee
provided that the denominator is nonzero, i.e.,
\be
A\neq \frac{5\lambda\mu^2-8\mu^3\pm\mu^2\sqrt{25\lambda^2+208\lambda\mu+64\mu^2}}{18\lambda}\;.
\ee
As a concrete example of this, we see that since, in flat space, $\Theta^2=0$ for only $A=\{\mu^2/3\;, 2\mu^3/(2\mu+3\lambda)\}$, it means that Eq. \eqref{s-fcg} can be satisfied for non-vanishing $\omega^a$ and $\Theta$ provided $A\neq\{\mu^2/3\;, 2\mu^3/(2\mu+3\lambda)\}$\;.

\item For flat space,  $\lambda=0$ and $A\neq\{\mu^2, \mu^2/3\}$. \\In this case, we can have non-vanishing $\Theta$ and $\omega^a$ provided that
\be
\Theta=\pm\sqrt{\frac{3A-\mu^2}{4\mu}}\;.
\ee
\end{enumerate}

\section{Conclusion}
In this work, we have shown that the inclusion of the original Chaplygin gas in the original GR gravitational field equations will lead to a possibility of a simultaneous rotation and expansion of the universe in the almost FLRW model for the case of linearized sheer-free perturbations.
Unlike previously known general relativistic results with pure dust, the Chaplygin-fluid model with simultaneously rotating and expanding solutions presents a closer connection to Newtonian gravity where such dynamics is allowed. We have shown that the results of our Chaplygin gas cases all reduce
to the well-known GR results of a pressure-free universe. 

Although the current investigation involved the simplest case of the cosmological Chaplygin gas models (the original Chaplygin gas), it is evident from our results that the more general cases will have a bigger set of possibilities violating the classic shear-free conjecture.
\section*{Acknowledgments}
AA \& ME acknowledge that this work is based on the research supported in part by the
National Research Foundation (NRF) of South Africa (grant numbers 112131 \& 116657, respectively).  AA also acknowledges the hospitality of the High Energy and Astroparticle Physics Group of the Department of Physics of Sultan Qaboos University, where most of this work was completed.  HN thankfully acknowledges the financial support provided by Science and Engineering Research Board (SERB), New Delhi,  during the course of this work through grant number EMR/2017/000339.


\begin{thebibliography}{99}

\expandafter\ifx\csname url\endcsname\relax
  \def\url#1{\texttt{#1}}\fi
\expandafter\ifx\csname urlprefix\endcsname\relax\def\urlprefix{URL }\fi
\providecommand{\bibinfo}[2]{#2}
\providecommand{\eprint}[2][]{\url{#2}}
%
\bibitem{godel52} K. G{\"o}del. Rotating universes in general relativity theory. In Proceedings of the International Congress of Mathematicians Edited by LM Graves et al., Cambridge, Mass., {\bf 1} 175, 1952.
\bibitem{ellis67} G. R. Ellis. Dynamics of pressure-free matter in general relativity. Journal of Mathematical Physics, {\bf 8} 1171, 1967.
\bibitem{nzioki2011shear} Nzioki, Anne Marie and Goswami, Rituparno and Dunsby, Peter KS and Ellis, George FR. Shear-free perturbations of Friedmann-Lema{\^\i}tre-Robertson-Walker universes. Physical Review D, {\bf 84} 124028 (12), 2011.
\bibitem{ellis2011} G. F. Ellis. Shear free solutions in general relativity theory. General Relativity and Gravitation, {\bf 43 }(12) 3253, 2011.
\bibitem{narlikar99} J. V. Narlikar. Spinning universes in {N}ewtonian cosmology. In On Einstein's Path, 319-327. Springer, 1999.
\bibitem{narlikar63} J. Narlikar. {N}ewtonian universes with shear and rotation. Monthly Notices of the Royal Astronomical Society, {\bf 126} 203, 1963.
\bibitem{abebe2011shear}  A. Abebe, R. Goswami, and P.K.S. Dunsby. Shear-free perturbations of $f(R)$ gravity. Physical Review D, {\bf 84} (12) 124027, 2011.
\bibitem{nandan15} I. Pahwa, H. Nandan, and U.D. Goswami. Shear Dynamics in Higher Dimensional FLRW Cosmology, Astrophysics and Space Science, {\bf  360} 58, 2015.
\bibitem{nandan09} A. Dasgupta, H. Nandan, and S. Kar. Kinematics of geodesic flows in stringy black hole backgrounds. Physical Review D, {\bf 79} 124004, 2009.
\bibitem{EllisCovariant} G.~F.~R.~Ellis \& H van Elst,
Cosmological Models, Carg\`{e}se Lectures 1998, in Theoretical
and Observational Cosmology, Dordrecht: 1, Kluwer,
1999.
\bibitem{chap1904}
\bibinfo{author}{Chaplygin, S.}
\newblock \bibinfo{title}{On gas jets, {S}ci. {M}em., {M}oscow {U}niv. {P}hys-{M}ath. {\bf 21} 1, 1904}.
\newblock \bibinfo{journal}{Trans. by M. Slud, Brown University}, \bibinfo{year}{1944}.
\bibitem{gorini2003}
\bibinfo{author}{Gorini, V.}, \bibinfo{author}{Kamenshchik, A.},
  \bibinfo{author}{Moschella, U.} \& \bibinfo{author}{Pasquier, V.}
\newblock \bibinfo{title}{The {C}haplygin gas as a model for dark energy}.
\newblock In \bibinfo{booktitle}{Proceedings of the MG10 Meeting held at
  Brazilian Center for Research in Physics (CBPF)}, \bibinfo{volume}{{\bf 20}} \bibinfo{pages}{26} , \bibinfo{organization}{World Scientific}, \bibinfo{year}{2003}.


\bibitem{bento02}
\bibinfo{author}{Bento, M.}, \bibinfo{author}{Bertolami, O.} \&
  \bibinfo{author}{Sen, A.}
\newblock \bibinfo{title}{Generalized {C}haplygin gas, accelerated expansion,
  and dark-energy-matter unification}.
\newblock \bibinfo{journal}{Physical Review D}
  \textbf{\bibinfo{volume}{66}}, \bibinfo{pages}{043507}, \bibinfo{year}{2002}.

\bibitem{bento03}
\bibinfo{author}{Bento, M.}, \bibinfo{author}{Bertolami, O.} \&
  \bibinfo{author}{Sen, A.}
\newblock \bibinfo{title}{Generalized {C}haplygin gas model: Dark
  energy-dark matter unification and {CMBR} constraints}.
\newblock \bibinfo{journal}{General Relativity and Gravitation}
  \textbf{\bibinfo{volume}{35}}, \bibinfo{pages}{2063}, \bibinfo{year}{2003}.

\bibitem{bilic02}
\bibinfo{author}{Bili{\'c}, N.}, \bibinfo{author}{Tupper, G.~B.} \&
  \bibinfo{author}{Viollier, R.~D.}
\newblock \bibinfo{title}{Unification of dark matter and dark energy: the
  inhomogeneous {C}haplygin gas}.
\newblock \bibinfo{journal}{Physics Letters B}
  \textbf{\bibinfo{volume}{535}}, \bibinfo{pages}{171}, \bibinfo{year}{2002}.

\bibitem{deb04}
\bibinfo{author}{Debnath, U.}, \bibinfo{author}{Banerjee, A.} \&
  \bibinfo{author}{Chakraborty, S.}
\newblock \bibinfo{title}{Role of modified {C}haplygin gas in accelerated
  universe}.
\newblock \bibinfo{journal}{Classical and Quantum Gravity}
  \textbf{\bibinfo{volume}{21}}, \bibinfo{pages}{5609}, \bibinfo{year}{2004}.

\bibitem{dev03}
\bibinfo{author}{Dev, A.}, \bibinfo{author}{Alcaniz, J.} \&
  \bibinfo{author}{Jain, D.}
\newblock \bibinfo{title}{Cosmological consequences of a {C}haplygin gas dark
  energy}.
\newblock \bibinfo{journal}{Physical Review D}
  \textbf{\bibinfo{volume}{67}}, \bibinfo{pages}{023515}, \bibinfo{year}{2003}.

\bibitem{gorini03}
\bibinfo{author}{Gorini, V.}, \bibinfo{author}{Kamenshchik, A.} \&
  \bibinfo{author}{Moschella, U.}
\newblock \bibinfo{title}{Can the {C}haplygin gas be a plausible model for dark
  energy?}
\newblock \bibinfo{journal}{Physical Review D}
  \textbf{\bibinfo{volume}{67}}, \bibinfo{pages}{063509}, \bibinfo{year}{2003}.
  \bibitem{kamen01}
\bibinfo{author}{Kamenshchik, A.}, \bibinfo{author}{Moschella, U.} \&
  \bibinfo{author}{Pasquier, V.}
\newblock \bibinfo{title}{An alternative to quintessence}.
\newblock \bibinfo{journal}{Physics Letters B}
  \textbf{\bibinfo{volume}{511}}, \bibinfo{pages}{265}, \bibinfo{year}{2001}.
  \bibitem{fabris02}
\bibinfo{author}{Fabris, J.~C.}, \bibinfo{author}{Gon{\c{c}}alves, S.~V.} \&
  \bibinfo{author}{de~Souza, P.~E.}
\newblock \bibinfo{title}{Letter: density perturbations in a universe dominated
  by the {C}haplygin gas}.
\newblock \bibinfo{journal}{General Relativity and Gravitation}
  \textbf{\bibinfo{volume}{34}}, \bibinfo{pages}{53}, \bibinfo{year}{2002}.

\bibitem{bean03}
\bibinfo{author}{Bean, R.} \& \bibinfo{author}{Dore, O.}
\newblock \bibinfo{title}{Are {C}haplygin gases serious contenders for the dark
  energy?}
\newblock \bibinfo{journal}{Physical Review D}
  \textbf{\bibinfo{volume}{68}} \bibinfo{pages}{023515}, \bibinfo{year}{2003}.
\bibitem{bouh15}
\bibinfo{author}{Bouhmadi-L{\'o}pez, M.}, \bibinfo{author}{Brilenkov, M.},
  \bibinfo{author}{Brilenkov, R.}, \bibinfo{author}{Morais, J.} \&
  \bibinfo{author}{Zhuk, A.}
\newblock \bibinfo{title}{Scalar perturbations in the late universe: viability
  of the {C}haplygin gas models}.
\newblock \bibinfo{journal}{Journal of Cosmology and Astroparticle
  Physics} \textbf{\bibinfo{volume}{2015}}, \bibinfo{pages}{037}, \bibinfo{year}{2015}.
\bibitem{tup}N. Bilic, G. B. Tupper, and R. D. Viollier, Unification of dark matter and dark energy: the inhomogeneous Chaplygin gas,  Physics Letters B, {\bf 535} (1), 17, 2002.
\bibitem{pas} A. Kamenshchik, U. Moschella, and V. Pasquier, “An alternative to quintessence, Physics Letters B, {\bf 511} (2) 265, 2001.
\bibitem{elmardi17} M. Elmardi and A. Abebe. Cosmological Chaplygin gas as modified gravity, Journal of Physics: Conference Series, {\bf 905} 012015, 2017.
\bibitem{barr87}
\bibinfo{author}{Barrow, J.~D.}
\newblock \bibinfo{title}{Deflationary universes with quadratic lagrangians}.
\newblock \bibinfo{journal}{Physics Letters B}
  \textbf{\bibinfo{volume}{183}}, \bibinfo{pages}{285}, \bibinfo{year}{1987}.

\bibitem{mak03}
\bibinfo{author}{Makler, M.}, \bibinfo{author}{de~Oliveira, S.~Q.} \&
  \bibinfo{author}{Waga, I.}
\newblock \bibinfo{title}{Constraints on the generalized {C}haplygin gas from
  supernovae observations}.
\newblock \bibinfo{journal}{Physics Letters B}
  \textbf{\bibinfo{volume}{555}}, \bibinfo{pages}{1}, \bibinfo{year}{2003}.

\bibitem{set07}
\bibinfo{author}{Setare, M.}
\newblock \bibinfo{title}{Interacting generalized chaplygin gas model in
  non-flat universe}.
\newblock \bibinfo{journal}{The European Physical Journal C}
  \textbf{\bibinfo{volume}{52}}, \bibinfo{pages}{689}, \bibinfo{year}{2007}.
  \bibitem{abebe13} A. Abebe. Beyond Concordance Cosmology, PhD thesis, University of Cape Town, 2013.
  \bibitem{ellis98} G. F. Ellis and H. van Elst. Cosmological models (Carg\`ese Lectures), arXiv [gr-qc] 9812046, 1998.
\end{thebibliography}
\end{document}